\documentclass[twocolumn,showpacs,preprintnumbers,amsmath,amssymb]{revtex4}

\usepackage{graphicx}
\usepackage{dcolumn}
\usepackage{bm}

\begin{document}
\title{A Mutual Attraction Model for Both Assortative and Disassortative Weighted Networks}
\author{Wen-Xu Wang$^{1}$}
\author{Bo Hu$^{1}$}
\email{hubo25@mail.ustc.edu.cn}
\author{Bing-Hong Wang$^{1}$}
\author{Gang Yan$^{2}$}
\affiliation{%
$^{1}$Nonlinear Science Center and Department of Modern Physics,
University of Science and Technology of China, Hefei, 230026, PR
China  \\
$^{2}$Department of Electronic Science and Technology,
University of Science and Technology of China, Hefei, 230026, PR
China
}%
\date{\today}

\begin{abstract}
In most networks, the connection between a pair of nodes is the
result of their mutual affinity and attachment. In this letter, we
will propose a Mutual Attraction Model to characterize weighted
evolving networks. By introducing the initial attractiveness $A$
and the general mechanism of mutual attraction (controlled by
parameter $m$), the model can naturally reproduce scale-free
distributions of degree, weight and strength, as found in many
real systems. Simulation results are in consistent with
theoretical predictions. Interestingly, we also obtain nontrivial
clustering coefficient $C$ and tunable degree assortativity $r$,
depending on $m$ and $A$. Our weighted model appears as the first
one that unifies the characterization of both assortative and
disassortative weighted networks.
\end{abstract}

\maketitle

The past few years have witnessed a great deal of interest from
physics community to understand and characterize the underlying
mechanisms that govern complex networks. Prototypical examples
cover as diverse as the Internet \cite{Internet}, the World-Wide
Web \cite{WWW}, the scientific collaboration networks (SCN)
\cite{CN1,CN2}, and world-wide airport networks
(WAN)\cite{air1,air2}. As a landmark, Barab\'asi and Albert (BA)
proposed their seminal model that introduces the linear
preferential linking to mimic the topological evolution of complex
networks \cite{BA}. However, networks are far from boolean
structure. The purely topological characterization will miss
important attributes often encountered in real systems. For
example, the amount of traffic characterizing the connections of
communication systems or large transport infrastructure is
fundamental for a full description of these networks \cite{top10}.
This thus calls for the use of weighted network representation,
which is often denoted by a weighted adjacency matrix with element
$w_{ij}$ represents the weight on the edge connecting vertices $i$
and $j$. In the case of undirected graphs, weights are symmetric
$w_{ij}=w_{ji}$, as this letter will focus on.  A natural
generalization of connectivity in the case of weighted networks is
the vertex strength described as
$s_{i}=\sum_{j\in\Gamma(i)}w_{ij}$, where the sum runs over the
set $\Gamma(i)$ of neighbors of node $i$. This quantity is a
natural measure of the importance or centrality of a vertex in the
network. Most recently, the access to more complete empirical data
and higher computation capability has allowed scientists to
consider the variation of the connection weights of many real
graphs. As confirmed by measurements, complex networks not only
exhibit a scale-free degree distribution $P(k)\thicksim
k^{-\gamma}$ with 2$\leq\gamma\leq$3 \cite{air1,air2}, but also
the power-law weight distribution $P(w)\sim w^{-\theta}$
\cite{ref1} and the strength distribution $P(s)\sim s^{-\alpha}$
\cite{air2}. Highly correlated with the degree, the strength
usually displays scale-free property $s\thicksim k^{\beta}$ with
$\beta\geq1$ \cite{air2, traffic-driven, empirical}. Motivated by
those findings, Alain Barrat \emph{et al.} presented a model (BBV
for short) to study the growth of weighted networks \cite{BBV}.
Controlled by a single parameter $\delta$, BBV model can produce
scale-free properties of degree, weight and strength. But its
disassortative property \cite{BBV, GBBV} (i.e. the hubs are
primarily connected to less connected nodes), as observed in real
technological and biological networks, can hardly give satisfying
interpretations to social networks like the SCN, where the hubs
are very likely to be linked together (assortative mixing).
Previous models as far as our knowledge can generate either
assortative networks \cite{vazquez, assortative, assortative2} or
disassortative ones \cite{BBV, GBBV, WWX, vazquez}, but rarely
both. Thus, some questions arises here: why are social networks
all assortative, while all biological and technological networks
opposite? Is there a generic explanation for the observed
incompatible patterns, or does it represents a feature that needs
to be addressed in each network individually? Our work may shed
some new light to these questions.

Former network models often impress on us such a network evolution
picture: pre-existing nodes are passively attached by newly added
node according to the preferential linking mechanism. This
scenario, however, lacks the other side of fact that old nodes
will choose the young at the same time. In addition, this
evolution picture also ignores the universal mutual attraction
between existing components, which leads to the creation and
reinforcement of connections. This idea has been partly reflected
in the studies of Dorogovtsev and Mendes (DM) \cite{DM} who
proposed a class of undirected and unweighted models where new
edges are added between old sites and existing edges can be
removed. In this letter, we will present a model for weighted
evolving networks that considers the topological evolution under
the general mechanism of mutual attraction between nodes. In
contrast with previous models where weights are assigned
statically \cite{ref2, ref3} or rearranged locally \cite{BBV,
GBBV}, our model allow weights to be widely updated. It can mimic
the reinforcement and creation of internal links as well as the
evolution of many infrastructure networks. Specifically, the model
can generate a diversity of scale-free quantities, nontrivial
clustering property, and tunable assortativity coefficient.
Therefore, one can easily find explanations to various real
networks by our microscopic mechanisms.

The model starts from an initial configuration of $N_0=m$ isolated
nodes with no connections between each other. At each time step, a
new isolated node $n$ is introduced into the system. Then every
existing node $i$ (including the newly-added one) selects $m$
other existing nodes for potential interaction with the
probability
\begin{equation}
\Pi_{i\rightarrow j}=\frac{s_j+A}{\sum_{k(\neq i)}(s_k+A)}.
\end{equation}To guarantee the isolated new node be chosen
by others the model requires $A\geq0$, and hence $A$ is called the
initial attractiveness \cite{DM2}. If two unconnected nodes are
mutually selected, then an internal connection is created between
them. If there already exists a connection between them, their
link is just strengthened by increasing weight $w_0=1$. Here, $m$
is the number of candidate nodes (of each site per step) to create
and strengthen connections, and the initial attractiveness $A$
governs the probability for ``young" nodes to get new links and
weights. After the weights have been updated, the growth process
is iterated by introducing a new vertex until the desired size of
the network is reached.

One can easily find appropriate interpretations to real networks
from our model mechanism. Take the SCN for example: the
collaboration of scientists require their common interest and
mutual acknowledgements. Unilateral effort does not promise
effective activity. Speaking by the model language, though the
nodes with low degree would like to connect to nodes with large
degree, the latter do not necessarily wish to be linked by former.
On the other hand, two scientists both with strong scientific
potentials (large strengths) and long collaborating history are
more likely to publish papers together. The above description of
our model could also satisfactorily explain the WAN where the edge
weigh denotes the relative magnitude of the traffic along a flight
line. During the evolution of WAN, the airlines are more likely to
open between metropolises that hold a high status in both economy
and politics (with large strengths). With the improvement of
economy and the expansion of population, the air traffic between
connected metropolises will increase much faster than that between
smaller cities. Due to their importance, there is an obvious need
for smaller cities to build new airports to connect those
metropolises. But the limit of energy and resources leads to the
fact each node can only afford a finite number of connections.
Therefore, they have to choose in front of the vertex pool.

\begin{figure}
\scalebox{0.80}[0.80]{\includegraphics{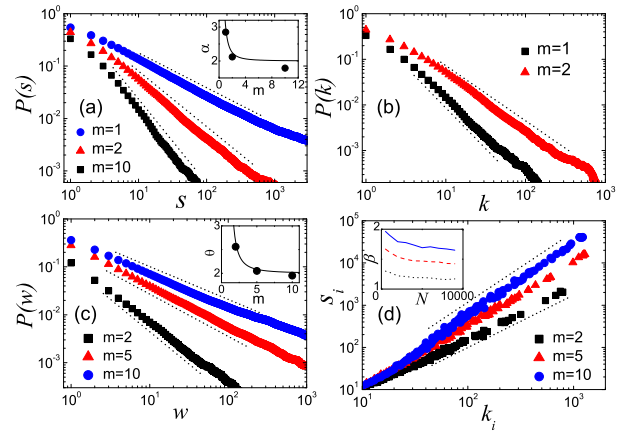}}
\caption{\label{fig:epsart} Numerical simulations by choosing
$A=1$. Data are averaged over 10 independent runs of network size
$N=8000$: (a) Cumulative probability strength distribution $P(s)$
with various values of $m$. Data are consistent with a power-law
behavior $P(s)\sim s^{-\alpha}$. The inset reports the values of
$\alpha$ obtained by data fitting (full circles) in comparison
with the theoretical prediction $\alpha=2+A/m^2$ (line). (b)
Cumulative probability degree distribution $P(k)$ with $m=1$ and
$m=2$. Data fitting confirms its scale-free property. (c)
Cumulative probability distribution of weight with different $m$,
in agreement with the power-law tail $P(w)\sim w^{-\theta}$. As
shown in its inset, the data fitting also gives values of $\theta$
(full circles) as predicted by analytical calculation (line). (d)
The average strength $s_i$ of nodes with connectivity $k_i$ for
different $m$. In the log-log scale, we observe the nontrivial
strength-degree correlation $s\sim k^{\beta}$, with the exponent
$\beta$ versus network size $N$ (see the inset).}
\end{figure}

The model time is measured with respect to the number of nodes
added to the graph, i.e. $t=N-N_0$, and the natural time scale of
the model dynamics is the network size $N$. Using the continuous
approximation, we can treat $k, w, s$ and the time $t$ as
continuous variables \cite{BA, BBV}. The time evolution of the
weights $w_{ij}$ can be computed analytically as follows:
\begin{eqnarray}
\frac{dw_{ij}}{dt}&=&m\frac{s_j+A}{\sum_{k(\neq i)}(s_k+A)}\times
m\frac{s_i+A}{\sum_{k(\neq j)}(s_k+A)} \nonumber\\
&\approx&\frac{m^2(s_i+A)(s_j+A)}{\sum_k(s_k+A)\sum_k(s_k+A)}.
\end{eqnarray}Hence, the strength $s_i(t)$ is updated by this
rate:
\begin{equation}
\frac{ds_i}{dt}=\sum_j\frac{dw_{ij}}{dt}\approx\frac{m^2(s_i+A)}{\sum_k(s_k+A)}=\frac{m^2(s_i+A)}{(m^2+A)t}.
\end{equation}The last expression is recovered by noticing
that
\begin{equation}
\sum_i(s_i+A)=\sum_is_i+At=\int_0^t\frac{d\sum_is_i}{dt}dt+At=(m^2+A)t.
\nonumber
\end{equation} From Eq. (3),
one can obtain the scaling of $s_i(t)$ versus $t$ as $s_i(t)\sim
t^{\lambda}$, which also implies the scale-free distribution of
strength $P(s)\sim s^{-\alpha}$ with the exponent \cite{BBV}
\begin{equation}
\alpha=1+\frac{1}{\lambda}=1+\frac{m^2+A}{m^2}=2+\frac{A}{m^2}.
\end{equation} One can also obtain the evolution
behaviors of weight and degree, and hence their power-law
distributions \cite{MAM}: $P(w)\sim w^{-\theta}$ with
\begin{equation}
\theta=2+\frac{2A}{m^2-A}
\end{equation}
and $P(k)\sim k^{-\gamma}$ with $\gamma\rightarrow2+A/m^2=\alpha$
as $t\rightarrow\infty$.

\begin{figure}
\scalebox{0.80}[0.80]{\includegraphics{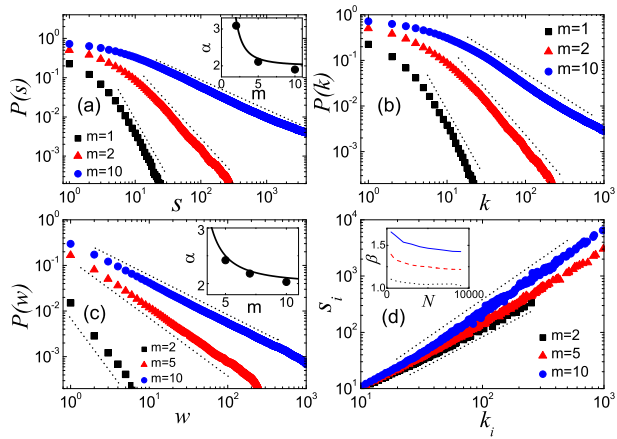}}
\caption{\label{fig:epsart} Numerical simulations by choosing
$A=5$. Data are averaged over 10 independent runs of network size
$N=8000$: (a) Cumulative probability strength distribution $P(s)$
with various values of $m$. Data are consistent with a power-law
behavior $P(s)\sim s^{-\alpha}$. The inset reports the values of
$\alpha$ obtained by data fitting (full circles) in comparison
with the theoretical prediction(line). (b) Cumulative probability
distribution of degree for different $m$. Data fitting confirms
its scale-free property: $P(k)\sim k^{-\gamma}$. (c) Cumulative
probability distribution of weight with different $m$, in
agreement with the power-law tail $P(w)\sim w^{-\theta}$. As shown
in its inset, the data fitting also gives values of $\theta$ (full
circles) as predicted analytically (line). (d) The average
strength $s_i$ of nodes with connectivity $k_i$ for different $m$.
In the log-log scale, we observe the nontrivial strength-degree
correlation $s\sim k^{\beta}$, with the exponent $\beta$ versus
network size $N$ (see the inset).}
\end{figure}

\begin{figure}
\scalebox{0.75}[0.75]{\includegraphics{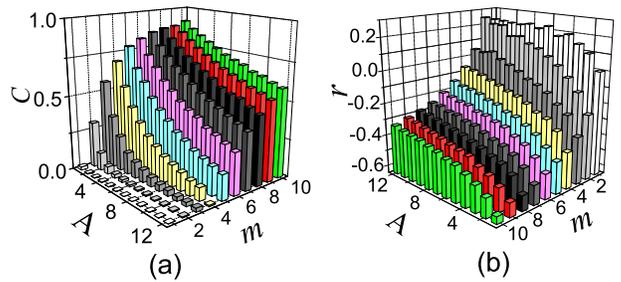}}
\caption{\label{fig:epsart} (a) Clustering coefficient $C$
depending on both $m$ and $A$ with network size $N=8000$. (b)
Degree assortativity $r$ depending on both $m$ and $A$ with
network size $N=8000$.}
\end{figure}

We performed numerical simulations of networks generated by
choosing different values of $A$ and $m$ and recovered the
theoretical predictions. We have also checked that the scale-free
properties of our model networks are almost independent of the
initial conditions. Fig. 1(a)-(d) report the probability
distributions of strength, weight and degree, as well as the
strength-degree correlation, fixed $A=1$ and tuned by $m$.
Specifically, Fig. 1(a) gives the probability distribution
$P(s)\sim s^{-\alpha}$, which is in good agreement with the
theoretical expression [Eq. 4]. Probability weight distribution
also recovers the power-law behavior $P(w)\sim w^{-\theta}$ [Fig.
1(b)] with $\theta$ as predicted analytically [Eq. 5]. Fig. 1(c)
shows the scale-free degree distribution $P(k)\sim k^{-\gamma}$
and Fig. 1(d) reports the average strength of vertices with degree
$k_i$, which displays a nontrivial power-law behavior $s\sim
k^{\beta}$ as confirmed by empirical measurements. The inset of
Fig. 1(d) indicates that the exponent $\beta$ decreases slowly
with the network size \cite{MAM}, which is noticeably different
from the linear correlation ($\beta=1$) as obtained in most
previous models. Again, Fig. 2(a)-(d) show the simulation results
by fixing $A=5$ and adjusting $m$. In comparison with Fig.
1(a)(b), the distributions of strength and degree for $A=5$ both
behave exponential corrections in the zone of low degree. This
phenomenon occurs at large $A$ and the exponential part are very
similar with the empirical findings in some social networks like
SCN \cite{Newman}. In the zone of large degree, however, we can
still observe the scale-free behavior which again recover the
theoretical exponent expressions. It is worth remarking ahead that
the model at large $A$ can generate the assortative property too,
which is seen in social networks. Thus, the introduction of $A$ is
important for our model to mimic social networks.

To better understand the degree correlations of our model
networks, we also studied the clustering coefficient $C$ (which
describes the statistic density of connected triples)
\cite{Newman} and degree assortativity $r$ \cite{mixing} depending
on the model parameters $A$ and $m$. As presented in Fig. 3(a),
$C$ for fixed $m$ monotonously decreases with $A$, and $C$ for
fixed $A$ monotonously increases with $m$. Generally, it can be
adjusted in the range ]0,1[. Obviously, the clustering property of
our model is tunable in a broad range by simultaneously varying
$m$ and $A$, which makes it more powerful in modelling real
networks. As shown in Fig. 3(b), degree assortativity $r$ for
fixed $m$, unlike the clustering case, increases with increasing
$A$; while $r$ for given $A$ decreases with $m$. For small $A$ and
large $m$, the model generates disassortative networks which can
best mimic technological networks like the Internet
\cite{Internet} and WAN or even biological networks. While at
large $A$ and small $m$, assortive weighted networks emerge and
can be used to model social graphs as the SCN. Actually, enhancing
the initial attractiveness $A$ will considerably increase the
chances for ``young" nodes to be linked and strengthened. As
low-degree nodes take the majority in the system, larger $A$ will
lead to the stronger affinity between ``young" vertices, and thus
they can link together more easily. This explains the origin of
assortative mixing in our model and may also shed some light on
the old open question: why social networks are different from
other networks in degree assortativity? Considering humans are
active elements, it is plausible that the components of social
networks possess considerable initial attractiveness (large $A$).
On the other end, as $m$ controls the interaction level of
internal connections, increasing $m$ will make the hubs become
busier and busier, as they have to be linked by more and more
``young" sites. It may explain why the disassortativity of the
model is increasingly sensitive to $m$. Combining these two
parameters together, the mutual attraction model integrates two
competitive ingredients that may be responsible for the mixing
difference in complex networks.

The universal mutual attraction between nodes and the existence of
initial node attractiveness are two important united ingredients
of our model in mimicking real weighted networks. The general
dynamics of node interaction proposed in this letter provides a
wide variety of scale-free behaviors, nontrivial clustering
coefficient and tunable degree assortativity. As far as our
knowledge, the weighted network model presented here appears as
the first one that can both mimic assortative and disassortative
networks under a unified evolution dynamics. Its obvious
simplicity and reproduced real-world variety allow more specific
mechanisms to be integrated into future modelling efforts. Above
all, the Mutual Attraction Model we presented here implies us the
possible and worthwhile efforts in exploring the unified
mechanisms behind various networks.

We thanks Yan-Bo Xie for his valuable comments and suggestions.
This work is funded by NNSFC under Grants No. 10472116 and No.
70271070.


\begin{thebibliography}{ref1}
\bibitem{Internet} R. Pastor-Satorras and A. Vespignani, {\it Evolution
and Structure of the Internet: A Statistical Physics Approach}
(Cambridge University Press, Cambridge, England, 2004).

\bibitem{WWW} R. Albert, H. Jeong, and A.-L. Barab\'asi, Nature \textbf{401}, 130 (1999).

\bibitem{CN1} M.E.J. Newman, Phys. Rev. E \textbf{64}, 016132 (2001).

\bibitem{CN2} A.-L. Barab\'asi, H. Jeong, Z. N\'eda. E. Ravasz, A.
Schubert, and T. Vicsek, Physica (Amsterdam) \textbf{311A}, 590
(2002).

\bibitem{air1} R. Guimera and L.A.N Amaral, Eur. Phys. J. B
\textbf{38}, 381 (2004).

\bibitem{air2} A. Barrat, M. Barth\'elemy, R. Pastor-Satorras, and
A. Vespignani, Proc. Natl. Acad. Sci. U.S.A. \textbf{101}, 3747
(2004).

\bibitem{BA} R. Albert and A.-L. Barab\'asi, Rev. Mod. Phys. \textbf{74}, 47 (2002).

\bibitem{top10} A virtual round tabel on ten leading questions for
network research can be found in the special issue on Applications
of Networks, edited by G. Caldarelli, A. Erzan, and A. Vespignani
[Eur. Phys. J. B \textbf{38}, 143 (2004)].

\bibitem{ref1} W. Li adn X. Cai, Phys. Rev. E \textbf{69}, 046106
(2004).

\bibitem{traffic-driven} K.-I. Goh, B. Kahng, and D. Kim,
cond-mat/0410078 (2004).

\bibitem{empirical} R. Pastor-Satorras, A. V\'azquez, and A.
Vespignani, Phys. Rev. Lett. \textbf{87}, 258701 (2001).

\bibitem{BBV} A. Barrat, M. Barth\'elemy, and A. Vespignani, Phys.
Rev. Lett. \textbf{92}, 228701 (2004).

\bibitem{GBBV} B. Hu, G. Yan, W.-X. Wang, and W. Chen, e-print
cond-mat/0505417.

\bibitem{vazquez} A. V\'azquez, Phys. Rev. E \textbf{67}, 056104
(2003).

\bibitem{assortative} R. Xulvi-Brunet and I. M. Sokolov, Phys. Rev. E \textbf{70}, 066102
(2004).

\bibitem{assortative2} M. Catanzaro, G. Caldarelli, and L. Pietronero, Phys.
Rev. E \textbf{70}, 037101 (2004).

\bibitem{WWX} W.-X. Wang, B.-H. Wang, B. Hu, G. Yan, and Q. Ou,
Phys. Rev. Lett. \textbf{94}, 188702 (2005).

\bibitem{DM} S.N. Dorogovtsev and J.F.F. Mendes, Europhys. Lett.
\textbf{52}, 33 (2000).

\bibitem{ref2} S.H. Yook, H. Jeong, A.-L. Barab\'asi, and Y. Tu, Phys. Rev. Lett. \textbf{86}, 5835 (2001).

\bibitem{ref3} D. Zheng, S. Trimper, B. Zheng, and P.M. Hui, Phys. Rev. E \textbf{67}, 040102 (2003).

\bibitem{DM2} S.N. Dorogovtsev and J.F.F. Mendes, Phys. Rev. Lett. 85, 4633 (2000).

\bibitem{MAM} B. Hu, Wen-Xu Wang \emph{et al.} (unpublished).

\bibitem{Newman} M. E. J. Newman, SIAM Review \textbf{45}, 167-256 (2003).

\bibitem{mixing} M. E. J. Newman, Phys. Rev. E \textbf{67}, 026126
(2003).
\end{thebibliography}
\end{document}